# Generative metamaterials based on large language models

**Authors**
Zhenyang Gao[1,2,3,4], Gengchen Zheng[1,2], Pengyuan Ren[1,2], Hongsong Wang[5], Kun Zhou[3], Minh-Son Pham[6], Yi Wu[1,2,7*], Yu Zou[4], Chu Lun Alex Leung[8,9], Yuanyuan Tian[3], Yang Lu[10], Haowei Wang[1,2,10], Hongze Wang[1,2,10,11*]

**Affiliations**
[1]State Key Labortory of Metal Matrix Composites, Shanghai Jiao Tong University, Shanghai, 200240, China
[2]School of Materials Science and Engineering, Shanghai Jiao Tong University, Shanghai, 200240, China
[3]School of Mechanical and Aerospace Engineering, Nanyang Technological University, Singapore, 639798, Singapore
[4]Department of Materials Science and Engineering, University of Toronto, Toronto, ON M5S 3E4, Canada
[5]Department of Computer Science and Engineering, Southeast University, Nanjing 210096, China
[6]Department of Materials, Imperial College London, London, SW7 2AZ, UK
[7]Institute of Alumics Materials, Shanghai Jiao Tong University (Anhui), Huaibei, 235000, China.
[8]Department of Mechanical Engineering, University College London, London, WC1E 7JE, UK
[9]Research Complex at Harwell, Harwell Campus, Oxfordshire, OX11 0FA, UK
[10] Department of Mechanical Engineering, the University of Hong Kong, Hong Kong, 999077, China.
[11]Shanghai Key Laboratory of Material Laser Processing and Modification, Shanghai, 200240, China.

*Corresponding author. Email: eagle51@sjtu.edu.cn, hz.wang@sjtu.edu.cn

**Abstract**

Mechanical metamaterials utilize intricate architectural designs to achieve advanced properties beyond those of their bulk counterparts. Existing metamaterial designs often rely on design inspirations and extensive experimental and numerical studies operated by design professionals, which can be time- and resource-consuming and limited in exploring the vast design space. Here, we transform metamaterial design by developing ChatMetamaterials based on large language models, a prompt-based generative metamaterial design engine capable of inventing architecture codes, and conducting reasoning-based diagnostics and evolution for complex metamaterial systems based on simple text prompts or hand-drawn sketches. This approach changes the way metamaterials are designed, and provides new opportunities for high-throughput metamaterial discovery.

## Introduction

The maturity of additive manufacturing presents new design potential through the creation of intrinsic architectures[1,2], enabling the development of architected metamaterials[3-6]. These metamaterials benefit from complex structural designs to achieve superior properties that are unattainable with conventional bulk materials. Such properties include extreme specific stiffness[7-9], tunable Poisson's ratio[10-12], ultrahigh energy absorption[13-16], smart mechanics[17-19], multi-stability[20-23], advanced optics[24], fracture resistance[25-27], irrational structures[5], or replicating of strengthening mechanisms in nature[28-37]. The design of new architectures often relies heavily on the material designers' intuition and expertise, limiting their wide adoption. To achieve reliable engineering of advanced metamaterial systems, artificial intelligence (AI), mostly machine learning (ML) [38-43], is increasingly used to guide the design of complex mechanical responses of metamaterials[44-60] based on carefully formatted data. However, the current use of AI remains on using AI as non-generative AI-assisted property prediction and optimization tool, which hinders the acceleration of new metamaterial development. In addition, human-developed structural design strategy, codes, or algorithms currently restrict AI-driven metamaterial invention, as AI designs remain confined within existing structural generation frameworks. The wide adoption of metamaterials not only necessitates a transformation of metamaterial design concepts demanding high-level expertise, and requires extensive cross-industry knowledge generalization capability of next-generation design engine.

Large language models (LLMs), a subset of generative artificial intelligence (AI) [61,62], can generate new content based on a new format of multimodal and polysemous text or images data, supporting the discovery and optimization of new materials[63-67], protein structures[68], and the next-generation engineering applications[69-71]. Recently, specialized large language models, particularly fine-tuned OpenAI GPT models[72,73], have been widely utilized to support research across various fields through AI-based scientific knowledge reasoning. Here, we transform the metamaterial design by introducing the prompt-based metamaterial design engine with in-depth reasoning capability named ChatMetamaterials (Fig. 1, Supplementary Movie 1, details of ChatMetamaterials system designs and fine-tuning application programming interfaces (APIs) in Methods, Supplementary Fig. 5, Supplementary Note 1-4). The unique capability of LLMs enables ChatMetamaterials to interpret multimodal prompts, invent architectural codes supporting multiphysics, perform reasoning-based property diagnostics, and evolve by comprehending experimental data enriched with deep human knowledge feedback (Supplementary Fig. 1). This distinctive ability not only fundamentally transforms communication within the design process (Fig. 2) but also speed up both the rapid invention of entirely new topologies (Fig. 3) and the fast in-depth exploration of extensive, complex topology spaces in existing systems (Supplementary Fig. 2).

## Results
*Prompt-based communications in metamaterial designs*

Rapid and intuitive generation of known metamaterials with prompt-defined functional properties (Fig. 2) is critical for lowering the design barriers, enabling flexible re-design and re-use of known architectures, and supporting rapid iteration in practical deployments of mechanical metamaterials. ChatMetamaterials addresses this through interpreting user prompts (textual descriptions or hand-drawn sketches) based on carefully fine-tuned GPT-4o APIs (details in Methods and Supplementary Note 4) and directing them to specialized metadata for given metamaterial design algorithms. We showcase this advancement through three types of fundamental tasks on a programmable topology with dynamical cell centroids (design and manufacturing details in Methods and Supplementary Note 5): (1) engineering unit cell properties (Fig. 2a); (2) recognizing and programming complex deformation patterns based on detailed user-provided descriptions (Fig. 2b); and (3) generating practical product designs (Fig. 2c), translating user sketches and text prompts into specific geometries, dimensions, thicknesses, and modulus distributions.

For fundamental mechanical properties, a hand-drawn Ashby chart specifying stress and energy absorption values for an "AI" logo was accurately translated into fabricated lattice architectures with maximum deviations of only 0.77 MPa in compressive strength and 12.7 mJ/(g/cm³) in specific energy absorption. Complex deformation pattern recognition capabilities were confirmed through experiments where generated metamaterials precisely matched deformation and failure sequences outlined in sketches, verified via digital image correlation (DIC) analysis (see Supplementary Fig. 6 and Supplementary Fig. 7 for details), with strain fields replicating targeted failure sequences at 15% global compressive strain. In practical product design, experimental measurements of AI-designed shoe soles validated ChatMetamaterials' accuracy in thickness and modulus distributions, achieving user-requested moduli values precisely, with measured stiffness values closely matching the prompt requirements (position A-D, moduli ranged between 8.09 and 20.15 MPa). To explore the robustness of interpretation across different prompt engineering strategies, a detailed one-to-many analysis is given (Supplementary Fig. 3 and Supplementary Fig. 8) to systematically evaluate the uncertainties inherent to varying levels of abstraction. The results not only provides prompt engineering guidelines for uses to develop metamaterials with different levels of inspiration freedom supported by AI, but also confirmed the precise fine-tuning capabilities of ChatMetamaterials based on explicit user feedback, thus validating its robust performance and adaptability to diverse user requirements.

The demonstrated capabilities highlight the practical potential of ChatMetamaterials in design of metamaterials. By accurately interpreting user-defined prompts, it minimizes reliance of metadata generation for a given metamaterial system on manual expertise, significantly reducing design costs through iterative cycles. This enables intuitive exploration of novel metamaterial architectures and facilitates broader accessibility for engineers and designers without specialized knowledge (see Supplementary Fig. 4 for using ChatMetamaterials for research exploration). The engineerable precision in replicating user-defined properties, deformation patterns, and product geometries confirms ChatMetamaterials' role as a robust tool to democratize complex metamaterial design processes, enhancing broader industrial applicability, and opening pathways for innovative structural discoveries.

*AI-inventions of new metamaterial topologies and codes*
The core of new metamaterial creation relies fundamentally on inventing new architectures and algorithms, tasks that traditionally required human experts to provide explicit design codes or structural frameworks. Such human-dependent constraints significantly limited the potential for fundamental AI-driven invention of entirely new metamaterial architectures. Here, we complete the final piece of the AI-driven metamaterial design puzzle by developing a metamaterial AI coder equipped with advanced reasoning capabilities. We fine-tuned a GPT-4.1 API using data from over 2,000 research papers related to metamaterials, code design guidelines, and design codes derived from our previous programmable metamaterial design publications in past five years (detailed fine-tuning methods in Methods and Supplementary Note 2). The resulting "ChatMetamaterials: Algorithm Coding" agent can autonomously develop new code, inventing topologies based on human prompts, alongside detailed property predictions, design intuition explanations, and deep scientific reasoning linked directly to the generated codes and architectures. We illustrate the reasoning-based code generation capabilities of ChatMetamaterials (Fig. 3) through its invention of over 200 new mechanical, thermal, and fluid metamaterials, accompanied by insightful knowledge explanations and property predictions for subsequent experimental validations.

Fig. 3a depicts the agent inventing and forecasting mechanical properties of more than 200 novel energy-absorbing metamaterials across 11 categories (F1–F11), bridging soft to stiff regimes (reasoning details available in Supplementary Note 3). Experimentally measured specific energy absorptions (SEAs) and specific compressive modulus closely match the property space anticipated by ChatMetamaterials across these categories (uncertainty data detailed in Supplementary Fig. 9a-b). Moreover, it autonomously coded a multifunctional, actively controllable aquatic robot inspired by water striders, incorporating freeform legs constructed from 26 distinct, 'ChatMetamaterials' generated functional meta-structures.

Fig. 3b demonstrates a thermal camouflage metamaterial designed for soft wearable devices, concealing portions of a human hand while selectively revealing an "AI" logo. The wearable camouflage effectively blocked human heat radiation, with the cloaking region showing a maximum temperature deviation of 0.9°C compared to the environment, while the "AI" logo region maintained a temperature of $36\pm0.5$°C (Supplementary Fig. 10), clearly distinguishable in infrared imaging . Alongside this camouflage material, ChatMetamaterials invented 10 additional metamaterials with adjustable thermal resistances. Experimental validations confirm that these metamaterials' temperatures increased from 35.6°C to 38.5°C after 120 seconds of exposure to a 40°C heating pad, matching the AI-predicted range of $35\pm2$°C to $40\pm2$°C (Supplementary Fig. 10). Fig. 3c presents a selectively impermeable porous metamaterial composed of AI-coded microfluidic units and various fluid-controlling cells devised by ChatMetamaterials. These innovative structures leverage specific dimensions related to micro-fluid knowledge, precisely controlling fluid propagation despite their porosity.

ChatMetamaterials' detailed reasoning processes offer scientific insights that bridge structural motifs and emergent mechanical properties. For example, the generation of soft, highly deformable metamaterials leverages architectural cues from neural microcoil entanglements, plant tendril networks, and lamellar interfaces observed in fish scales (see Supplementary Note 3). These compliant frameworks feature hierarchical interconnections, high topological disorder, and sliding interfaces, collectively enabling pronounced energy dissipation, stress delocalization, and damage tolerance under large deformations. As the architecture evolves toward stiffer, more load-bearing regimes, ChatMetamaterials draws upon inspirations such as hierarchical skeletal cages, cuboidal shells with engineered porosity, and bio-inspired grass cluster arrays (see Supplementary Note 3). These topologies utilize thickened load paths, multi-axial strut arrangements, and structural gradients to direct stress, suppress catastrophic failure, and achieve superior stiffness-to-density ratios. For instance, the grass cluster-inspired architecture employs densely splayed fiber bundles radiating from a robust base, enabling rapid load redistribution and localized energy absorption, characteristic of advanced rigid energy-dissipating systems. By tightly coupling machine-driven logical synthesis with multi-scale bio-inspiration, ChatMetamaterials systematically alters the discovery of rational metamaterial architectures, ensuring that each structural feature is explicitly linked to target mechanical outcomes across the soft–stiff property spectrum.

Similarly, the thermal cloaking cells harness mechanisms such as compartmentalized air trapping, curved insulating membranes, and tortuous heat pathways to achieve selective heat blocking, effectively suppressing both conductive and convective transfer at the microscale (see Supplementary Note 3). In contrast, thermal camouflage cells are engineered with continuous, highly printable frameworks that strategically minimize thermal resistance, allowing for controlled heat transfer and integration within multifunctional devices. For fluid control, the microfluidic cells employ physical phenomena such as capillary exclusion, directional viscous resistance, and micro-scale Venturi gating (see Supplementary Note 3). Here, fluid entry is blocked or precisely guided through sub-millimeter voids and geometric constrictions, with critical dimensions and topologies determined through AI-driven generative design.

*Reasoning-based metamaterial diagnostics and evolution*
Previously, in-depth understanding of metamaterial design knowledge, abstract structure-performance insights, and toughening mechanisms traditionally relies solely on human experts. Conventional AI-driven methods typically offer only data-driven numerical or metadata predictions for structural optimization without the capability to interpret and reason about complex design concepts or deep functional phenomena. This limitation substantially slows the advancement of new metamaterial development, and the rapid AI-driven understanding of complex existing metamaterial systems.

To overcome this bottleneck, we report an AI reasoning-based diagnostics and evolutionary framework within ChatMetamaterials, moving beyond simple numerical property predictions. ChatMetamaterials performs human-like diagnostics on metamaterial designs, swiftly evolving its knowledge to explore extensive design spaces and invent new metamaterial configurations tailored to specific functional goals. In tasks requiring sophisticated understanding and detailed research knowledge, ChatMetamaterials demonstrates its ability to conduct human-like reasoning and evolves its understanding iteratively (Fig. 4a, see Supplementary Fig. 1 for a sample reasoning process of diagnostics and evolution, also see Supplementary Note 6 for research topic exploration). To benchmark the capability of learning a complicated metamaterial system, ChatMetamaterials was employed to invent advanced energy-absorbing, lightweight, and soft metamaterials based on a given topology (See Supplementary Note 5 for base topology design details). It provided comprehensive research backgrounds, insights into design strategies, and clear guidelines for interpreting and processing research data in a human-like manner (prompt details provided in Supplementary Table 1). Through five iterative rounds with 10 samples per iteration, we experimentally validated the model's learning capacity from relatively small datasets (detailed benchmarking designs in Supplementary Note 8; metadata specifics in Supplementary Fig. 11). Each round's experimental data with knowledge package, including normalized specific energy absorption (SEA) data, design details, description of experimental observations, and human suggestions, were fed back into ChatMetamaterials, enhancing its internal knowledge database (Fig. 6a). Results (Fig. 6b) indicated that normalized SEA values improved significantly from $1.33\times10^4$ to $2.96\times10^5$ mJ·MPa$^{-1}$·g$^{-1}$·cm³ across iterations, achieving performance up to 17–20 times superior compared to GAN- and FCNN-based methods (Fig. 6b-c). We also quantify the evolution rate ($v_{\text{norm SEA}}$) and prediction error ($\alpha_{\text{norm SEA}}$) as defined by the reasoning-based framework in Methods. Our results reveal an average evolution rate that is 8x and 29x higher five rounds compared to FCNN and GAN, respectively, and a prediction error that is reduced by 81% relative to FCNN, averaged across five evolution rounds (Supplementary Fig. 7c and Supplementary Fig. 12).

In addition, we demonstrate a new form of metamaterial evolution beyond typical metadata or design parameter optimizations—algorithmic code evolution driven by in-depth knowledge. Fig. 3d illustrates real-time interaction with the "ChatMetamaterials: Algorithm Coding" agent, showcasing revisions to existing design codes. An initial grass-inspired metamaterial was refined upon request to feature more realistic, curved structures, followed by an added requirement for selective thermal resistance. The agent successfully updated the code and generated designs fulfilling both revisions. This new structure evolution capability provides enhanced fundamental coding design freedom of architecture deviation, introducing new structural features and design principles beyond current parameter-driven optimizations. The reported reasoning-based AI approach provides an innovative pathway for deeper understanding and more rapid development of complex metamaterial systems, marking a critical step forward in the intelligent and flexible creation of advanced materials.

**Discussion**

In summary, we present ChatMetamaterials, the agent-based design engine built upon LLMs, capable of interpreting multimodal design prompts, inventing new metamaterial design codes, and conducting scientific diagnostics and evolution informed by knowledge feedback. ChatMetamaterials translates diverse prompts into design metadata, enabling the creation of metamaterials and associated products with programmable properties and establishing a new paradigm for design communication. The algorithm-coding agent within ChatMetamaterials autonomously generates new structural codes and provides scientific rationales underlying complex architectural designs, advancing metamaterial design. Beyond property prediction, the model offers reasoning-based diagnostics, elucidating design intuitions, functional mechanisms, and structure–property relationships for each architecture. ChatMetamaterials also incorporates abstract human feedback and observational insights, enabling dynamic revision of both design metadata and foundational coding algorithms, thereby supporting the iterative enhancement of functionalities and structural features. ChatMetamaterials changes the traditional way of existing metamaterial design, while also serving as a beneficial tool for the rapid discovery and development of previously unexplored materials and structures in engineering applications.

**Materials and Methods**

*Architecture design*

The architectures presented in this study were generated using python based on the RhinoPythonScript syntax. These designs were created either by applying ChatMetamaterials-generated metadata to pre-defined design algorithms or directly using new design codes invented by ChatMetamaterials. ChatMetamaterials was carefully fine-tuned from OpenAI's GPT-4o and GPT-4.1 models to support both metadata generation for existing topologies and the invention of entirely new topological design codes (see Supplementary Note 1, Supplementary Note 2, and Supplementary Note 4 for details). The printable models were exported as .stl files using Rhino 6 and subsequently prepared for fabrication using Formlabs® PreForm software.

*Reasoning-based diagnostics and evolution theory*

This work introduces reasoning-based diagnostics and evolution methodologies to support the rapid development and scientific understanding of new metamaterials. A comprehensive theoretical framework is established to describe the end-to-end pipeline of prompt interpretation, design, diagnostics, and iterative evolution in metamaterial discovery.

Within this paradigm, the ChatMetamaterials engine ($\mathcal{C}$) assimilates human-provided prompts to generate new metadata ($\mathcal{D}$) or invent new architecture codes ($\mathcal{A}$). These representations are processed by $\mathcal{C}$ to yield specific quantitative property predictions, and diagnostic reasoning packages:

$$\{\mathcal{P}_1^0, \mathcal{P}_2^0, \ldots, \mathcal{P}_n^0, \mathcal{K}_1^0, \mathcal{K}_2^0, \ldots, \mathcal{K}_m^0\} = \mathcal{C}_0(\mathcal{A}_0, \mathcal{D}_0) \qquad (1)$$

where $\mathcal{P}_n^0$ and $\mathcal{K}_m^0$ denote the $n^{th}$ property and $m^{th}$ knowledge packages, respectively, in the initial ($0^{th}$) evolution round (see Supplementary Fig. 1 for examples of $\mathcal{K}$ and $\mathcal{P}$ representing fracture knowledge and numerical fracture data predictions).

Subsequently, these design packages ($\mathcal{A}$, $\mathcal{D}$) are fabricated and experimentally validated or subjected to computational studies, yielding empirical property measurements and knowledge insights:

$$\{\mathcal{P}_{e,1}^0, \mathcal{P}_{e,2}^0, \ldots, \mathcal{P}_{e,n}^0, \mathcal{K}_{e+h,1}^0, \mathcal{K}_{e+h,2}^0, \ldots, \mathcal{K}_{e+h,m}^0\} = \mathcal{O}(\mathcal{A}_0, \mathcal{D}_0) \quad (2)$$

where $\mathcal{P}_{e,n}^0$ indicates the nth property determined via experiment or simulation, and $\mathcal{K}_{e+h,m}^0$ represents human knowledge packages synthesized from expert interpretation and experimental observations. Here, $\mathcal{O}$ signifies the combined process of experimental/numerical evaluation and human knowledge integration.

Leveraging both AI-derived diagnostics and accumulated human expertise, informed design suggestions are formulated by the human agent ($\mathcal{H}$):

$$\{\mathcal{S}_{h,1}^0, \mathcal{S}_{h,2}^0, \ldots, \mathcal{S}_{h,k}^0\} = \mathcal{H}(\mathcal{K}_{e+h,1}^0, \ldots, \mathcal{K}_{e+h,m}^0, \mathcal{K}_1^0, \ldots, \mathcal{K}_m^0) \quad (3)$$

where $\mathcal{S}_{h,k}^0$ denotes the kth suggestion based on integrated human knowledge. These insights, together with the empirical property data, are reintroduced into ChatMetamaterials to generate an evolved suite of design codes, updated metadata, and an enriched model state:

$$\{\mathcal{A}_1, \mathcal{D}_1, \mathcal{C}_1\} = \mathcal{C}_0(\mathcal{S}_{h,1}^0, \ldots, \mathcal{S}_{h,k}^0, \mathcal{P}_{e,1}^0, \ldots, \mathcal{P}_{e,n}^0, \mathcal{K}_{e+h,1}^0, \ldots, \mathcal{K}_{e+h,m}^0) \quad (4)$$

where $\mathcal{A}_1$, $\mathcal{D}_1$, and $\mathcal{C}_1$ correspond to the evolved design code, metadata, and ChatMetamaterials model, respectively.

To quantitatively characterize the evolution from round i–1 to round i, we define the following indices:

$$\alpha_{\mathcal{P}_x} = |\mathcal{P}_{e,x}^i - \mathcal{P}_x^i| / \mathcal{P}_{e,x}^i \quad (5)$$

$$\nu_{\mathcal{P}_x} = |\mathcal{P}_{e,x}^i - \mathcal{P}_{e,x}^{i-1}| / \mathcal{P}_{e,x}^{i-1} \quad (6)$$

where $\alpha_{\mathcal{P}_x}$ and $\nu_{\mathcal{P}_x}$ denote the prediction error and evolution rate, respectively, for property index $\mathcal{P}_x$ in the $i^{th}$ evolution round.

*Finite-element analysis*

The simulation models were generated using RhinoPythonScript in Rhino 6 and imported as beam elements into Abaqus CAE[74] for FEA calculations. Supplementary Fig. 13 illustrates the FEA configurations used to study the architected metamaterials[75], where two rigid plates at the top and bottom planes of the cells were used to apply compressive strain. The bottom plate was fixed, while a uniform displacement of 4 mm was applied to the top plate with a displacement rate identical to that in the experiment. The beam elements of the cells and the compressive plates were meshed with B32 and S4R elements, respectively. Reaction forces and displacements during compression were recorded. Based on the displacement, reaction force, topology, and coordinates of the centroid location, a training dataset was constructed.

*Materials and experiments*

The raw material for all structures studied in this paper was selected to be Formlabs® durable resin for comparison purposes, although the presented method is not limited to this raw material. The properties of the raw material were evaluated according to the

ASTM D638[76] standard with type-IV specimens on a ZwickRoell Z100 universal tensile test machine (Supplementary Fig. 14a). The resulting stress–strain curves are provided in Supplementary Fig. 14b. The raw material exhibited a modulus of 775.7 MPa, an ultimate tensile stress of 30.1 MPa, and a fracture strain of 41.5%. All samples were printed using a Formlabs® Form 3 3D printer, ultrasonically washed in a pure ethanol solution for 5 minutes, cured with ultraviolet light at 60 °C for 60 minutes, and stored in a dark environment for 24 hours prior to mechanical tests (Supplementary Fig. 15). Supplementary Fig. 16 illustrates the compressive test configuration of the metamaterials. All compressive tests were performed on the ZwickRoell Z100 universal tensile test machine at a displacement rate of 1 mm/minute to ensure quasi-static load conditions, with samples compressed to 80%. The load–displacement data were recorded during the tests, while the following mechanical relations are applied to calculate the stress $\sigma$, strain $\varepsilon$, and specific energy absorption $SEA$, respectively:

$$\sigma = \frac{F}{A}, \tag{7}$$

$$\varepsilon = \frac{\delta}{S}, \tag{8}$$

$$SEA = \int_0^{\delta_d} F/\rho \, d\delta, \tag{9}$$

where $F$ is the compressive load, $A$ is the cross-sectional area, $\delta$ is compressive strain, $S$ is the height of the metamaterial sample, $\rho$ is the density, and $\delta_d$ is the displacement of the densification.

*Digital image correlation analysis*
DIC experiments were performed to calculate the strain field during the compression of the samples. All samples for DIC analysis were prepared with a white background and scattered black spot coatings, where the black spots served as reference points during the DIC calculation. The data were recorded using a camera with a 25-mm lens, capturing images at a resolution of 4096 × 3000 pixels. The step dimensions were 1300 × 1400 pixels with a minimum unit size of 5 × 5 pixels during DIC calculations. Strain field calculations were performed using correlated solutions generated by the VIC-2D 7 commercial software.

**Acknowledgments**

**Funding:**
National Key Research and Development Program of China (No. 2023YFB3712001)
National Natural Science Foundation of China (523B2048)
National Natural Science Foundation of China (52075327 and 52004160)
Shanghai Sailing Program (20YF1419200)
Natural Science Foundation of Shanghai (20ZR1427500)
SJTU Global Strategic Partnership Fund (2023 SJTU-CORNELL)
The University Synergy Innovation Program of Anhui Province (GXXT-2022-086)
The Innovation Foundation of Commercial Aircraft Manufacturing Engineering Center of China (No. 3-0410300-031)
Research Grants Council of the Hong Kong Special Administrative Region, China (RFS2021-1S05)
Hong Kong RGC general research fund (#11200623)
RGC Hong Kong under the CRF project (C7074-23GF)
CLAL is grateful for the support from the UKRI – EPSRC grants (references: EP/R511638/1, EP/W006774/1, EP/P006566/1, EP/W003333/1, EP/V061798/1 and EP/W037483/1) and the IPG Photonics/Royal Academy of Engineering Senior Research Fellowship in SEARCH (ref: RCSRF2324-18-71)
The authors also acknowledge Chen Huan for assisting DIC experiments, Shenzhen Victeknix Technology Co., Ltd for thermal measurements, thank Yifeng Dong, Yujia Tian, Asker Jarlov, Devesh Kripalani, Weiming Ji, Shubo Gao, Chenyang Zhu in Prof. Zhou's group, and Juzheng Chen in Prof. Lu's group for assisting the paper revison.

**Author contributions:**
Z.G. conceptualized this study. Z.G., H.Z.W. and M.S.P. contributed to the methodology design of the study. Z.G., G.C.Z., and P.Y.R. conducted simulations. Z.G. and G.C.Z. operated experiments. Z.G. constructed training prompts. Z.G. performed data analysis, interpretation, visualization, implemented the ChatMetamaterials, and wrote the manuscript. Z.G., Y.T., H.Z.W., Y.Z., C.L.A.L.,

M.S.P., K.Z., Y.L., Y.W., and H.S.W. substantially revised the work. H.Z.W., K.Z., Y.W., Y.L., H.W.W. administered the project and supervised the study of this paper. Z.G., H.Z.W., Y.W. Y.L., Y.Z. conducted the funding acquisition for this work.

**Competing interests:** Authors declare that they have no competing interests.

**Data and materials availability:** All data related to this study are available in the main text or the supplementary materials.

**Figures and Tables**

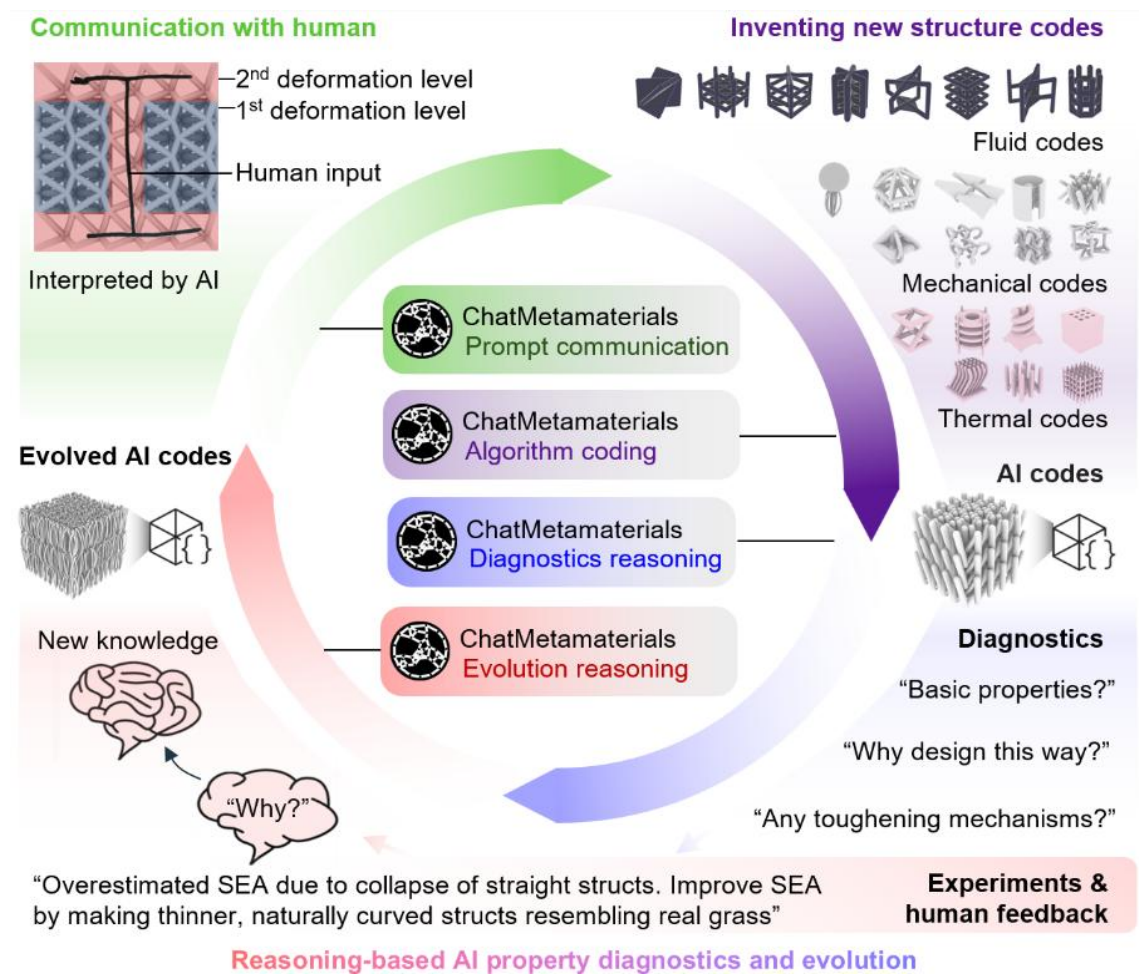

**Fig. 1. Metamaterial intelligent design cycle driven by large language models.** This includes metamaterial prompt-based communication (**Fig. 2**), code and architecture generation (**Fig. 3**), reasoning-driven property prediction and diagnostics (**Supplementary Fig. 1**), and knowledge-guided evolution (**Fig. 4**), enabled by ChatMetamaterials. Here the SEA is specific energy absorption, AI represents artificial intelligence.

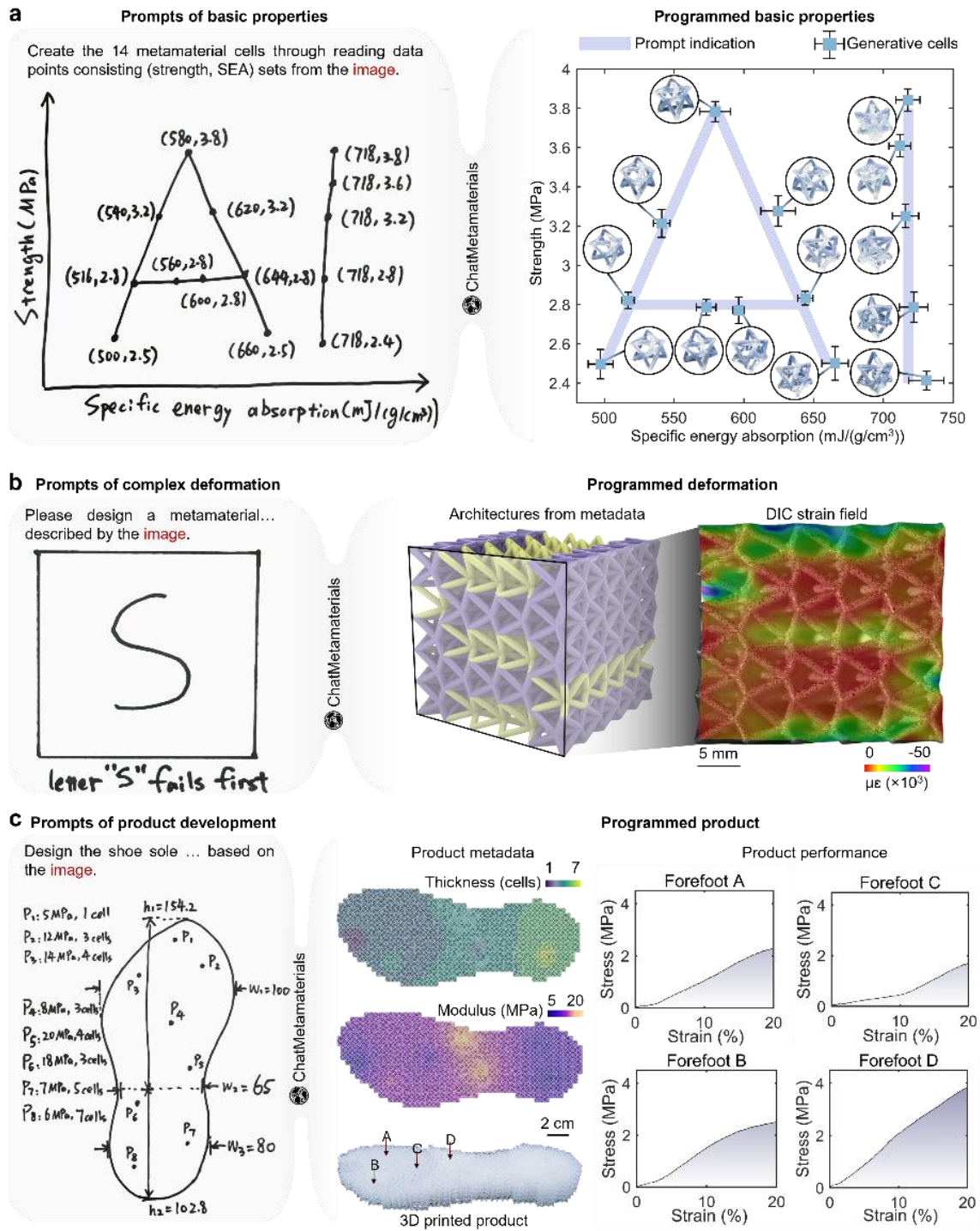

**Fig. 2. Prompt-based design of ChatMetamaterials. (a)** Unit cell designs and properties (strength and SEA-specific energy absorption) interpreted by ChatMetamaterials based on a hand-drawn Ashby data sketch (also see Supplementary Fig. 17 for stress-strain curves matching the prompted stress levels). **(b)** Translating complex deformation response patterns drawings to metamaterial architecture metadata, and digital image correlation (DIC) analysis of generated metamaterials at 15% compressive strain, respectively. **(c)** Prompt-based shoe sole product demo designed. The product metadata is designed based on a product prompt with hand-drawn image

illustrating a customer design idea for a shoe sole, detailing the specifications to describe dimensions and properties, where $w_1$, $w_2$, and $w_3$ represent the width of the forefoot, arch, and hindfoot, respectively; $h_1$ and $h_2$ indicate the length from the toe to arch and from the arch to the foot end, respectively; $p_1$ to $p_8$ denote points along the foot, from the toe to the right and left forefoot, forefoot-to-arch transition, right and left arch, and right and left hindfoot. (b) The thickness map, elastic modulus map, and a fabricated shoe sole created by ChatMetamaterials based on the product prompt. The experimental stress-strain curves represent the responses of cells at A-D positions of the forefoot, respectively. See Supplementary Fig. 8 and Supplementary Fig. 18 for prompt engineering strategies controlling the one-to-many uncertainty of material and product designs, and see Supplementary Fig. 6 for detailed DIC validation data measuring deformation at isolated regions.

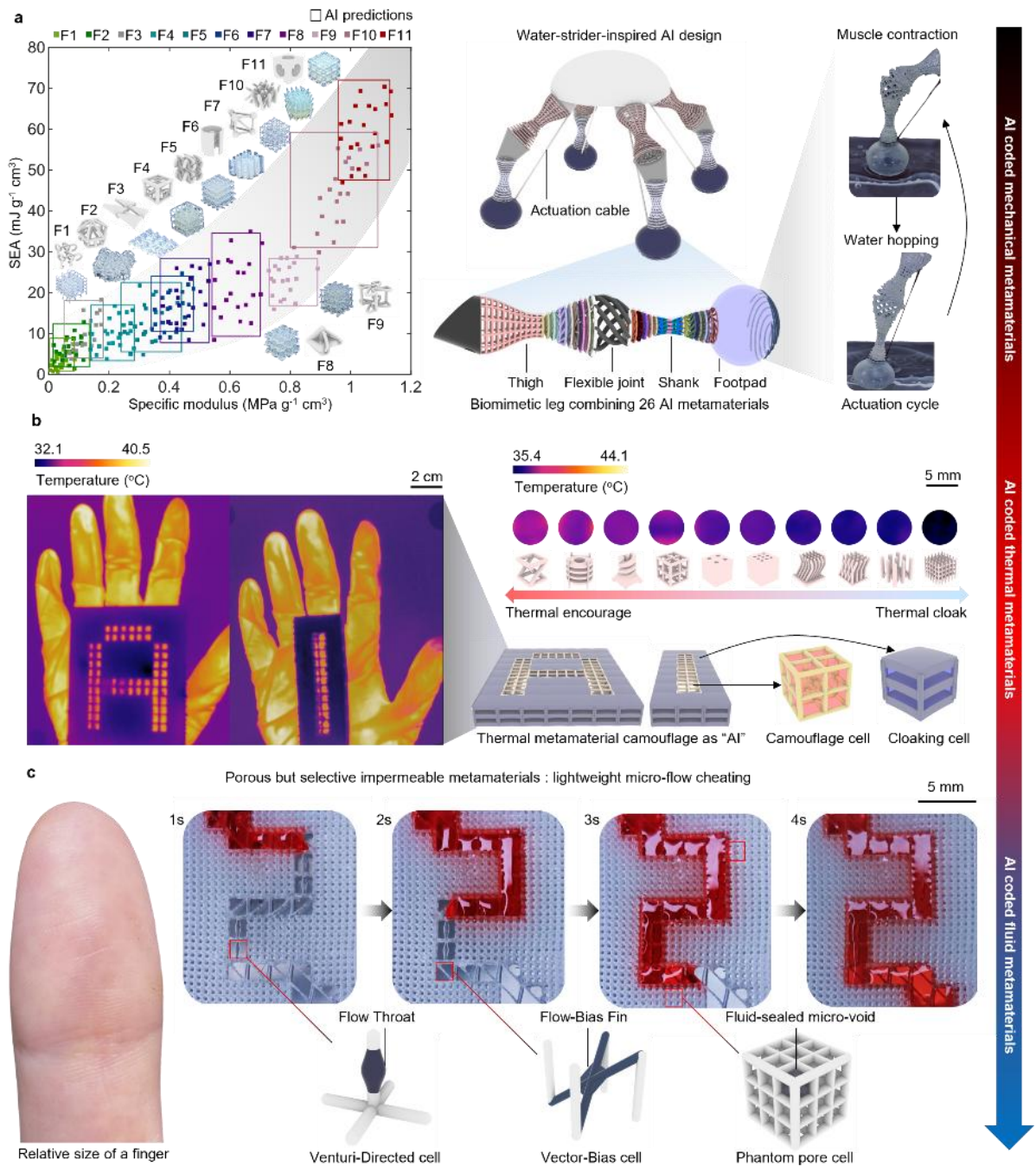

**Fig. 3. New mechanical, thermal, and fluid-control metamaterials and functional component code inventions.** (a) ChatMetamaterials codes 220 types of mechanical metamaterials across 11 broad categories, along with a multifunctional water-strider-inspired robot integrating 26 AI-designed metamaterial types. (b) 10+ types of thermal metamaterials—ranging from thermally conductive to cloaking—are generated, including a soft wearable thermal camouflage material that displays "AI" under infrared. Thermal images were captured using a VICTEKNIX VSC 600 infrared camera. (c) A porous metamaterial capable of selectively blocking and guiding fluid flow, along with various flow-guiding microcells coded by ChatMetamaterials.

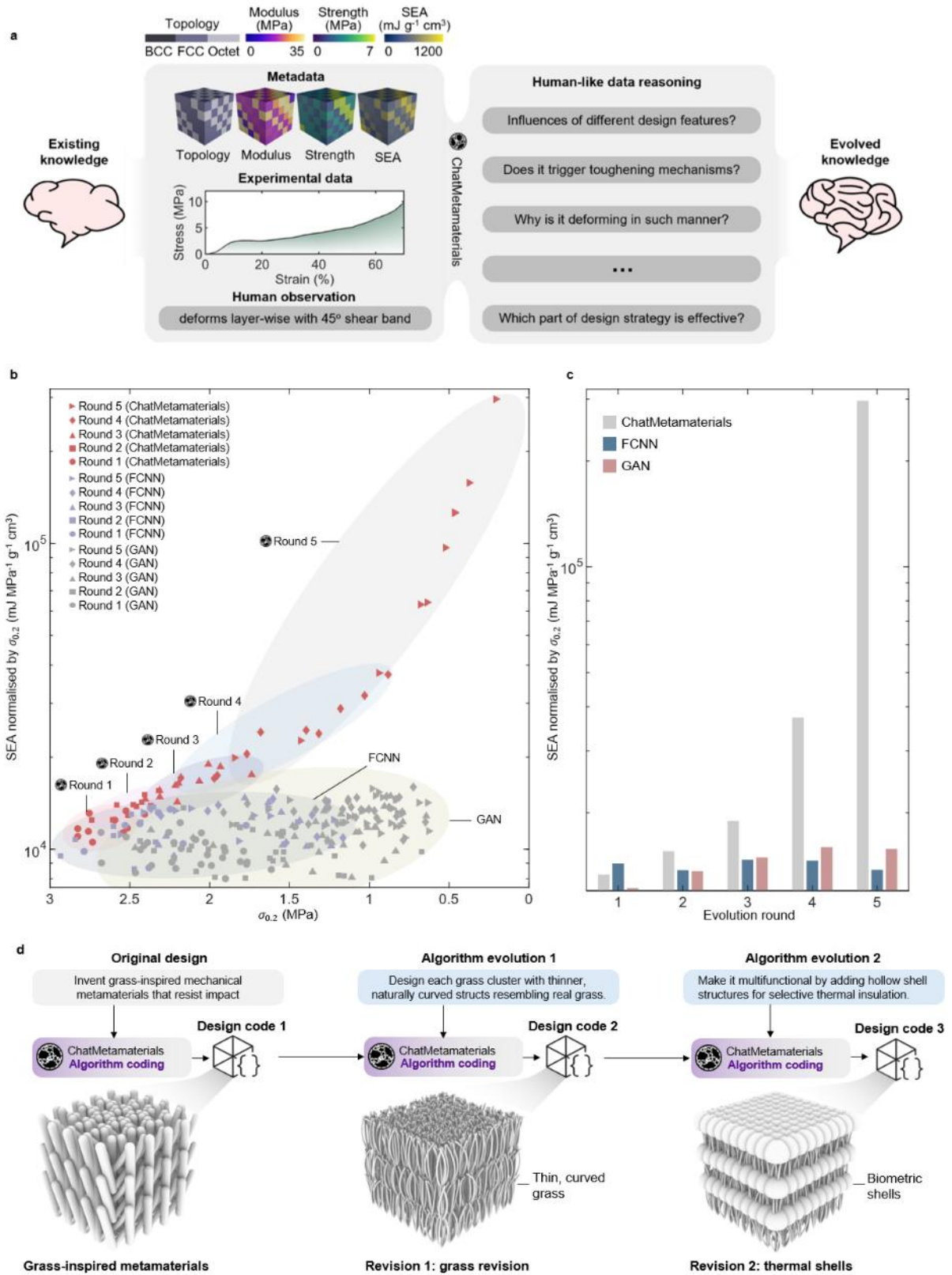

**Fig. 4. Knowledge-based metadata and code evolutions of metamaterials. (a)** Process of evolution and human-like learning process of research knowledge and metadata. The model is instructed by prompts in Supplementary Table 1 to conduct human-like reasoning. The metadata includes topology, modulus, strength, and specific

energy absorption (SEA) of metamaterials that have infinite design potentials and are described by 500 topological parameters (refer to Appeal Fig. 4 for typical reasoning process by ChatMetamaterials). **(b)** Ashby chart comparing the normalized SEAs, the initial stress responses at 20% strain ($\sigma_{0.2}$) evolved by ChatMetamaterials, FCNN, and GAN based design methods. **(c)** The maximum normalized SEAs at different evolution rounds by ChatMetamaterials, fully connected neural network (FCNN), and generative adversarial network (GAN) based design methods. **(d)** Evolution process of metamaterial structural codes based on human's knowledge feedback.